\documentclass[prd,twocolumn,noshowpacs,preprintnumbers,amsmath,amssymb,nofootinbib]{revtex4}
\usepackage{bm,calrsfs}
\def\a{\alpha}

\def\l{\lambda}

\def\m{\mu}
\def\n{\nu}

\def\ve{\varepsilon}
\def\x{\xi}

\def\be{\begin{equation}}
\def\ee{\end{equation}}
\def\beq{\begin{eqnarray}}
\def\eeq{\end{eqnarray}}

\def\ca{{\cal A}}

\def\cd{{\cal D}}

\def\ct{{\cal T}}

\def\cw{{\cal W}}

\begin{document}

\title{General solutions of the Wess-Zumino consistency condition for the Weyl anomalies}

\author{Nicolas Boulanger}
\altaffiliation{Charg\'e de recherches du F.N.R.S. (Belgium).}
\affiliation{\\ Universit\'e de Mons-Hainaut, Acad\'emie Wallonie-Bruxelles,\\
M\'ecanique et Gravitation, Avenue du Champ de Mars 6, B-7000 Mons, Belgium}

\begin{abstract} 
The general solutions of the Wess-Zumino consistency condition
for the conformal (or Weyl, or trace) anomalies are derived. The solutions
are obtained, in arbitrary dimensions,  
by explicitly computing the cohomology of the corresponding Becchi-Rouet-Stora-Tyutin 
differential in the space of integrated local functions at ghost number unity. 
This provides a purely algebraic, regularization-independent classification of 
the Weyl anomalies in arbitrary dimensions. The so-called type\,-A anomaly is shown to satisfy
a non-trivial descent of equations, similarly to the non-Abelian chiral anomaly
in Yang-Mills theory. 
\end{abstract}

\maketitle

\section{Introduction}

The Weyl (or conformal, or trace) anomalies have been discovered about 30 years 
ago~\cite{Capper:1974ic,Deser:1976yx} and still occupy a central position in theoretical physics, 
partly because of their important r\^oles within the AdS/CFT correspondance 
and their many applications in cosmology, particle physics, 
higher-dimensional conformal field theory, supergravity and strings.  
The body of work devoted to this subject is, therefore, considerable. 
A very non-exhaustive list of references can be found, e.g., in  \cite{Birrell:1982ix,Fradkin:1983tg,Duff:1993wm,Osborn:1993cr,Aharony:1999ti,Bastianelli:2006rx}.

The central equations which determine the candidate anomalies in quantum field theory 
are the Wess-Zumino (WZ) consistency conditions~\cite{Wess:1971yu}. 
By using these conditions, the general structure of all the know anomalies 
\emph{except the conformal ones} has been determined by purely algebraic methods 
featuring descent equations \`a la Stora-Zumino~\cite{Stora:1976kd,Zumino:1983ew}. 
We refer to the book~\cite{Bertlmann:1996xk} for a pedagogical review and
many references on the subject of anomalies in quantum field theory, 
while the works~\cite{Barnich:1995ap,Barnich:2000zw} 
contain and review the most general results for Einstein-Yang-Mills 
and Yang-Mills gauge theories, in the presence of antifields.  

As is well-known, the determination of the general 
solution of the WZ consistency conditions 
boils down to the computation of the cohomology of 
the corresponding BRST differential \cite{BRST} in the 
space of local functionals with ghost number one.
The cohomological formulation for the determination of the conformal anomalies 
was initiated in the pioneering works~\cite{Bonora:1983ff,Bonora:1985cq}, 
with results up to spacetime dimension~$n=6\,$.  
The authors of these works found that the Weyl anomalies comprise 
(i) the integral over spacetime of $\sigma$, the Weyl parameter, times the 
Euler density of the manifold, plus 
(ii) terms that are given by (the integral of) 
$\sigma$ times strictly Weyl-invariant scalar densities. 
Some of the terms from (ii) can be trivially obtained from contractions of 
products of the conformally invariant Weyl tensor,  
while the others are more complicated and involve covariant 
derivatives of the Riemann tensor. 
The same general structure was postulated~\cite{Bonora:1985cq} in higher (even) dimensions. 

These important cohomological results in dimensions $n=4$ and $n=6$ were obtained
by listing all the possible terms on the basis of dimensionality
and diffeomorphism invariance and by inserting them into the WZ 
consistency condition. The structure of the four-dimensional conformal anomalies 
was rederived later \cite{Cappelli:1988vw,Osborn:1991gm}, using the WZ conditions. 
Still, no systematic pattern emerged for the general structure of the Weyl anomalies 
in dimension~$n\,$. 
 
Such results appeared somewhat later, in~\cite{Deser:1993yx}.
By applying dimensional regularization on the effective gravitational action generated 
by a conformally invariant matter system, the authors of~\cite{Deser:1993yx} 
could confirm the structure found in~\cite{Bonora:1985cq}. 
The Euler term from class (i) was called ``type\,-A Weyl anomaly'', 
while the terms of (ii) were called ``type\,-B anomalies''. 
Very interestingly, from the structure of the poles in the effective action, 
it was observed~\cite{Deser:1993yx} that the type\,-A anomaly 
appeared in a similar way to the non-Abelian chiral anomaly in Yang-Mills gauge theory. 
That the type\,-A anomaly should arise via some ``descent identity'' was therefore suggested. 
Subsequently, this suggestion was taken as a work hypothesis in~\cite{Karakhanian:1994yd}.  
More recently, in the holographic context of the AdS/CFT correspondence where the
computation of the Weyl anomaly plays an important 
r\^ole~\cite{Witten:1998qj,Henningson:1998gx,Graham:1999pm}, 
some cohomological considerations have been applied~\cite{Imbimbo:1999bj,Manvelyan:2001pv} 
that confirm the structure found in~\cite{Bonora:1985cq,Deser:1993yx} and highlight the 
similarities between the type\,-A and the non-Abelian chiral anomalies. 

{}From these considerations, it appears that a purely algebraic understanding 
of the general structure of the Weyl anomalies, in arbitrary dimensions $n$ and 
\emph{independently} of the AdS/CFT correspondence or of any regularization scheme, 
is indeed most desirable and needed. 
It is also remarkable that, despite the enormous literature on the Weyl anomalies, 
these have not yet received the general algebraic treatment \`a la Stora-Zumino
that all the other known anomalies enjoy.
It is the purpose of the present paper to fill this gap, providing explicit proofs. 

More precisely, following the antifield-independent approach as in~\cite{Bertlmann:1996xk} 
and using the powerful cohomological tools reviewed in~\cite{Barnich:2000zw}, 
we solve the Wess-Zumino consistency condition for the Weyl anomaly 
in arbitrary dimensions $n\,$. We demonstrate that the type\,-A anomaly is the \textit{unique} 
solution associated with a non-trivial descent, whereas the type\,-B anomalies are given 
by trivial descents and can be computed by using the systematic, algebraic method 
of~\cite{Boulanger:2004zf,Boulanger:2004eh}. 
We do not resort to dimensional analysis and that the spacetime dimension $n$ must be even 
derives from consistency, it is not an assumption. 
These results are essentially obtained along the cohomological lines 
of~\cite{Brandt:1996mh,Brandt:1996au,Barkallil:2002fp} 
and crucially rely on preliminary results 
given in~\cite{Boulanger:2004eh}. 
They imply the uniqueness of the known conformal anomalies and solve a question 
posed in~\cite{Deser:1993yx} concerning the similitudes between the type\,-A anomaly and the 
non-Abelian chiral anomalies in Yang-Mills theories. 
[However, the precise expression for the non-trivial descent giving the 
type\,-A anomaly shows noticeable differences compared with the chiral anomaly.]

Incidentally, note also that our results provide a purely algebraic proof of the 
conjecture of differential geometry studied recently in~\cite{Alexakis:2005ft}\footnote{H. Baum 
is thanked for having pointed out these works to us.}. This is yet another instance 
of the rich interplay between the study of anomalies in theoretical physics 
and mathematics.

\section{Cohomological setting}
\label{sec:WeylAnomalies}

In a theory that is classically diffeomorphism and Weyl invariant, the associated BRST
differential is $s=s_{\!_D}+s_{\!_W}$, where $s_{\!_D}$ is 
the BRST differential corresponding to the diffeomorphisms and  
$s_{\!_W}$ corresponds to the Weyl transformations. 
As in \cite{Bonora:1985cq}, we consider the purely gravitational part of the 
cohomological problem, where the spacetime metric $g_{\mu\nu}$ is an external
classical field.   
Apart from the (invertible) metric, the other fields are the diffeomorphisms 
ghosts $\xi^{\mu}$ and the Weyl ghost $\omega\,$, with ghost number 
$gh(\xi^{\mu})=gh(\omega)=1\,$. 
Spacetime indices are denoted by Greek letters and run over the values $0,1,\ldots,n-1\,$. 
Flat, tangent space indices are denoted by Latin letters. 
The action of the BRST differential $s$ on the fields 
$\Phi^A=\{g_{\mu\nu},\xi^{\mu},\omega\}$ is
\begin{eqnarray}
s_{\!_D} g_{\mu\nu} &=& \xi^{\rho}\partial_{\rho}g_{\mu\nu}+\partial_{\mu}\xi^{\rho}g_{\rho\nu}
+\partial_{\nu}\xi^{\rho}g_{\mu\rho}\,,
\label{sdg} \\
s_{\!_W} g_{\mu\nu} &=& 2\,\omega \,g_{\mu\nu}\,,
\label{swg} \\
s_{\!_D} \xi^{\mu} &=& \xi^{\rho}\partial_{\rho}\xi^{\mu}\,,
\label{sdx} \\
s_{\!_D} \omega &=& \xi^{\rho}\partial_{\rho}\omega\,,
\quad s_{\!_W} \xi^{\mu} = 0 = s_{\!_W} \omega\,. 
\label{sdw} 	
\end{eqnarray}    
The anomalies $a_1^{n}$ are given by the solutions of 
the WZ consistency conditions
\begin{eqnarray}
	s a_1^n + d\, b_2^{n-1} = 0\,, ~\quad\quad~a_1^n \neq  s p_0^n + d\, q_1^{n-1}\,,
\label{WZbis}
\end{eqnarray}
where superscripts denote the form degree whereas subscripts indicate the ghost number. 
All the cochains $a_1^n$, $b_2^{n-1}$, $p_0^n$ and $q_1^{n-1}$ are local forms
and $d$ is the total exterior derivative. 
A local $p\,$-form $b^p$ depends on the fields $\Phi^A$ 
and their derivatives up to some finite (but otherwise unspecified) order, 
which is denoted by  
$b^p=\frac{1}{p!}\,d x^{\m_1}\ldots d x^{\m_p}\,b_{\m_1\ldots\m_p}(x,[\Phi^A])\,$. 

Since we are seeking Weyl anomalies, the ghost degree of $a^n_1$ is carried 
entirely by (a derivative of)~$\omega\,$. 
Decomposing the WZ consistency conditions (\ref{WZbis}) with respect to the 
Weyl-ghost degree, one finds
\begin{eqnarray}
	s_{\!_D} a^n_1 + d\, b^{n-1}_{2} &=& 0\,,
  \label{coho1}
	\\
	s_{\!_W} a^n_1 + d\, c^{n-1}_{2} &=& 0\,,\quad 
	a^n_1\neq s_{\!_W} p_0^{n} + d\, f_{1}^{n-1}\,,
	\label{coho2} \\
	&&\hspace*{1.6cm}s_{\!_D} p_0^n + d\, h_{1}^{n-1}=0\,.
	\label{coho3}
\end{eqnarray}
In words, we have to compute the cohomology $H^{1,n}(s_{\!_W}\vert d)$ 
of the Weyl BRST differential $s_{\!_W}$ modulo total derivatives, 
in the space of diffeomorphism-invariant local $n$-forms.
As a matter of fact, an important result of~\cite{Bonora:1985cq}
is that it is always possible, by adding a local Bardeen-Zumino 
counterterm to the action, 
to shift away the pure diffeomorphism part of the candidate anomaly $a_1^n$, 
leaving only the pure Weyl part of $a_1^n\,$.
This is consistent with the fact that it is always possible to
ensure diffeomorphism invariance throughout the process of regularization, 
at the price of losing Weyl invariance upon quantization. Actually, 
this can be taken as a definition of the Weyl anomaly.  

Before attacking the problem (\ref{coho1})--(\ref{coho3}), 
it is useful to reformulate the equations 
for the computation of $H^{1,n}(s\vert d)\,$ in slightly different terms. 
One can perform the Stora trick which consists in uniting the differentials 
$s=s_{\!_D}+s_{\!_W}$ and $d$ into a single differential $\tilde{s}=s + d\,$. 
Then, the WZ consistency condition (\ref{WZbis}) 
and its descent are encapsulated in 
\begin{eqnarray}
	\tilde{s} \,{\alpha} = 0\,, \quad {\alpha}\neq \tilde{s}\,{\zeta}+constant
\label{WZ}
\end{eqnarray}
for the local total forms ${\alpha}$ and ${\zeta}$ of total degrees $G=n+1$ and 
$G=n\,$. 
Local total forms are by definition formal sums of local forms with different form degrees 
and ghost numbers, ${\alpha}=\sum_{p=0}^n a^p_{G-p}\,$, the total degree
being simply the sum of the form degree and the ghost number.   
As proved in \cite{Brandt:1996mh}, the cohomology of ${s}$ in the space of local 
functionals (integrals of local $n$-forms) and at ghost number $g$ is locally 
isomorphic to the cohomology of $\tilde{s}$ in the space of local total forms
at total degree~$G=g+n\,$.   
Furthermore, the cohomological problem can be restricted, locally, 
to the $\tilde{s}$-cohomology on local total forms belonging to a subspace 
${\cal{W}}\,$ of the space of local total forms~\cite{Brandt:1996mh}:
\begin{eqnarray}
	&\tilde{s}\,\alpha({\cal{W}}) = 0\,,\quad 
	\alpha({\cal{W}})\neq \tilde{s}\,\zeta({\cal{W}})+constant\,,&
\label{cohoproblem1} \\
	&\quad totdeg (\alpha)=n+g\,, \quad totdeg (\zeta)=n+g-1\,.&
\nonumber	
\end{eqnarray}
The subspace ${\cal{W}}$, closed under the action of $\tilde{s}$, 
is given by local total forms depending on so-called tensor fields 
$\{ {\cal{T}}^i\}$ at total degree zero and on so-called generalized connections 
$\{\widetilde{C}^N\}$ at total degree unity. 
The latter decompose into a part with ghost number one and form degree zero 
plus a part having ghost number zero but
form degree unity: $\widetilde{C}^{N}=\widehat{C}^{N}+\ca^{N}\,$. 
For a purely gravitational theory in metric formulation,  
invariant under diffeomorphisms and Weyl transformations, 
the space ${\cal{W}}$ was found in~\cite{Boulanger:2004eh}.

The solution of the problem (\ref{WZ}) will thus have the form   
\begin{eqnarray}
 \a({\cal{W}})=\widetilde{C}^{N_1}\ldots\widetilde{C}^{N_n}\widetilde{C}^{N_{n+1}}
 \,a_{N_1\ldots N_{n+1}}({\cal{T}})
 \nonumber 	
\end{eqnarray}
where the anomalies are given (up to an unessential constant coefficient) 
by the top form-degree component of the local total form $\alpha({\cal{W}})$:
\begin{eqnarray} 
a^n_1={\cal{A}}^{N_1}\ldots{\cal{A}}^{N_n}\widehat{C}^{N_{n+1}}\,a_{N_1\ldots N_{n+1}}
({\cal{T}})\,. 	
\nonumber
\end{eqnarray} 

Now, we are ready to attack the system (\ref{coho1})---(\ref{coho3}).  
This is done by solving (\ref{cohoproblem1}) at total degree $G=n+1$  
with $\tilde{s}$ replaced by $\tilde{s}_{\!_W}={s}_{\!_W}+d$ and
taking the equations (\ref{coho1}), (\ref{coho3}) into account. 
These last two equations tell us that cocycles and 
coboundaries of $\tilde{s}_{\!_W}$ must be diffeomorphism-invariant. 
It is important to specify the space in which one computes the anomaly. 
Without any restriction of this kind, we would have the triviality of 
all the Weyl anomaly candidates $a_1^n=\omega f({\cal{T}})d^nx$ where 
$f(\ct)$ is a Weyl\,-invariant scalar density. 
Indeed, $\omega f({\cal{T}})d^nx = $
$\tilde{s}_{\!_W} [f({\cal{T}})d^nx\frac{1}{n}\,ln(\sqrt{-g}) ]\,$. 
However, the local form $p_0^n=\frac{1}{n}\,ln(\sqrt{-g})f({\cal{T}})d^nx$ 
is forbidden because it fails to obey the condition (\ref{coho3}).   
%
\section{Solution of the Wess-Zumino consistency condition }
%
To reiterate, we must look for $\tilde{s}_{\!_D}$-invariant 
$(n+1)$-local total forms $\alpha({\cal{W}})$ satisfying
\begin{eqnarray}
	&\tilde{s}_{\!_W}\alpha({\cal{W}})=0\,,
	\quad\alpha({\cal{W}})\neq \tilde{s}_{\!_W}\zeta({\cal{W}})+constant\,,&
  \label{cohoproblemweyl}
\end{eqnarray}
where $\zeta({\cal{W}})$ must be $\tilde{s}_{\!_D}$-invariant. 
The solution will take the general form
\begin{eqnarray}
\alpha({\cal{W}})=2\omega\;\tilde{C}^{N_1}\ldots\tilde{C}^{N_n}\;
a_{N_1\ldots N_{n}}({\cal{T}})\,.&
	\label{ansatz}
\end{eqnarray}
Before continuing with the solution of the WZ consistency condition for
the Weyl anomalies, we must spend some time in order to explain the
various symbols that appear in the above equation (\ref{ansatz}). 
In the same process, we will display the gauge covariant algebra
associated with the BRST transformations (\ref{sdg})--(\ref{sdw}) and
relate it to the conformal algebra $\mathfrak{so}(n,2)$, in the flat
space limit.   
 
The space ${\cal{T}}$ of tensor fields is generated by the (invertible) 
metric $g_{\mu\nu}$ together with the so-called $W$-tensors $\{W_{\Omega_i}\}$, $i\in\mathbb{N}$~\cite{Boulanger:2004eh}.  
It is only necessary to recall here that the $W$-tensors are tensors under general coordinate
transformations and transform under $s_{\!_W}$ according to 
$s_{\!_W} W_{\Omega_i}=\omega_{\alpha} \mathbf{\Gamma}^{\alpha}W_{\Omega_i}\,$, where
$\omega_{\alpha}=\partial_{\alpha}\omega\,$ and the $n$ generators $\mathbf{\Gamma}^{\alpha}\,$ 
$(0 \leqslant \alpha \leqslant n-1)\,$ act only on the $W$-tensors. 
These tensors are built recursively with the help of the formula $W_{\Omega_k}=(\nabla_{\alpha_k}+K_{\beta\alpha_k}\mathbf{\Gamma}^{\beta})W_{\Omega_{k-1}}={\cal{D}}_{\alpha_k}W_{\Omega_{k-1}}\,$, 
where $K_{\alpha\beta} = $
$\frac{1}{n-2}\,\Big({\cal{R}}_{\alpha\beta}-\frac{1}{2(n-1)}\,g_{\alpha\beta}{\cal{R}} \Big)\,$ and $W_{\Omega_0}=W^{\mu}_{~\;\nu\rho\sigma}$ is the conformally invariant
Weyl tensor. 
The symbol $\nabla$ denotes the usual torsion-free metric-compatible covariant 
differential associated with the Christoffel symbols $\Gamma_{~\,\nu\rho}^{\mu}\,$,   
while ${\cal{R}}_{\alpha\beta}=R^{\mu}_{~\,\alpha\mu\beta}\,$ is the Ricci tensor
with $R^{\mu}_{~\;\nu\rho\sigma}=\partial_{\rho}\Gamma_{~\,\nu\sigma}^{\mu}+\ldots$
the Riemann tensor. 
The scalar curvature is given by ${\cal{R}}=g^{\alpha\beta}{\cal{R}}_{\alpha\beta}\,$.

\noindent The Weyl tensor can be written as
\begin{eqnarray}
W^{\mu}_{~\;\nu\rho\sigma} = R^{\mu}_{~\;\nu\rho\sigma}
-2\left(\delta^{\mu}_{\,[\rho}K_{\sigma]\nu}-g_{\nu[\rho}K_{\sigma]}^{~\;\mu} \right)\,,	
\end{eqnarray}
where curved (square) brackets denote strength-one complete (anti)symmetrization. 

The following notation is useful and explains the meaning of the superindices 
$\Omega_i$, $i\in\mathbb{N}$: 
\begin{eqnarray}
 W_{\Omega_0}&=& W^{\mu}_{~\;\nu\rho\sigma} \;,\;\;
 W_{\Omega_1} = {\cal{D}}_{\alpha_1}W_{\Omega_0} 
 = {\cal{D}}_{\alpha_1}W^{\mu}_{~\;\nu\rho\sigma}\;,~\ldots
 \nonumber \\
 W_{\Omega_k}&=& {\cal{D}}_{\alpha_k}W_{\Omega_{k-1}} = 
 {\cal{D}}_{\alpha_k}{\cal{D}}_{\alpha_{k-1}}\ldots 
 {\cal{D}}_{\alpha_2}{\cal{D}}_{\alpha_1}W^{\mu}_{~\;\nu\rho\sigma}\,,
 \nonumber 
\end{eqnarray}
where $\cd$ is the Weyl-covariant derivative 
as introduced\footnote{V. W\"unsch informed us that such a construction 
had been obtained previously, see e.g.~\cite{GW} and references therein. 
Similar constructions and other references can be found in~\cite{Gover}. 
Apparently, all those works lead back to the ones of T. Y. Thomas~\cite{Thomas}.} 
in~\cite{Boulanger:2004eh}.

In the latter work we introduced and operator that counts the number of metric tensors
appearing in a given expression. An inverse metric brings a minus-one 
contribution. 
Explicitly, 
	$\Delta^{ex}_g = g_{\mu\nu}\frac{\partial}{\partial g_{\mu\nu}}\,$.
For example, $\Delta^{ex}_g (g_{\alpha\beta}g^{\gamma\delta})=0\,$
and
$\Delta^{ex}_g(g^{\gamma\sigma}g^{\lambda\nu}W_{\Omega_k})$
$=-2(g^{\gamma\sigma}g^{\lambda\nu}W_{\Omega_k})\,$. 
By definition, the operator $\Delta^{ex}_g$ gives zero when 
applied on the $W$-tensors $\{W_{\Omega_i}\,,~i\in{\mathbb{N}}\}$ and 
on the generalized connections $\{\tilde{C}^N\}\,$. 
Then, denoting\footnote{Notation is slightly changed as compared
with~\cite{Boulanger:2004eh}. In passing, we also correct a couple of
typos present therein.} by ${\Delta^{\mu}{}_{\nu}}$ 
the generators of $GL(n)$-transformations of world indices 
acting on a type\,-$(1,1)$ tensor $T_{\alpha}^{\beta}$ as
${\Delta^{\mu}{}_{\nu}} T_{\alpha}^{\beta}=\delta_{\alpha}^{\mu}T_{\nu}^{\beta}-\delta_{\nu}^{\beta}T_{\alpha}^{\mu}\,$,
the gauge covariant algebra $\cal{G}$ generated by 
$\{\Delta_N\}=\{\Delta^{ex}_g \,, {\cal{D}}_{\nu}\,,{\Delta^{\mu}{}_{\nu}}\,,\mathbf{\Gamma}^{\alpha} \}\,$
reads~\cite{Boulanger:2004eh}
\begin{eqnarray}
	{[}{\Delta^{\mu}{}_{\nu}},\mathbf{\Gamma}^{\alpha}{]} &=& -\delta_{\nu}^{\alpha}
	\mathbf{\Gamma}^{\mu}\,,\quad
	{[}{\Delta^{\mu}{}_{\nu}},{\cal{D}}_{\alpha}{]} = \delta_{\alpha}^{\mu}{\cal{D}}_{\nu}\,,
\label{gcov1} \\
	{[}{\Delta^{\rho}{}_{\mu}},{\Delta^{\sigma}{}_{\nu}}{]} &=& 
  \delta_{\nu}^{\rho}{\Delta^{\sigma}{}_{\mu}} - 
  \delta_{\mu}^{\sigma}{\Delta^{\rho}{}_{\nu}}\,,~~
	{[}\mathbf{\Gamma}^{\alpha},\mathbf{\Gamma}^{\beta}{]} = 0 \,,\quad  
\label{gcov2} \\
{[}{\cal{D}}_{\beta},\mathbf{\Gamma}^{\alpha}{]} &=& 
 {\cal{P}}^{\nu\alpha}_{\beta\mu}{\Delta^{\mu}{}_{\nu}} 
 - \delta^{\alpha}_{\beta}\Delta^{ex}_g\,,
\label{gcov3} \\
  {[}{\cal{D}}_{\rho},{\cal{D}}_{\sigma}{]} &=& 
  - W^{\mu}_{~\,\nu\rho\sigma}{\Delta^{\nu}{}_{\mu}} - 
  {C}_{\alpha\rho\sigma}\,\mathbf{\Gamma}^{\alpha}\,,
\label{algebra3}
\end{eqnarray} 
where ${C}_{\alpha\mu\nu} = {2}\,\nabla_{[\nu}K_{\mu]\alpha}\,$ is the Cotton tensor
and 
${\cal{P}}^{\nu\alpha}_{\beta\mu}=(-g^{\nu\alpha}g_{\beta\mu}
+\delta^{\nu}_{\beta}\delta^{\alpha}_{\mu}+\delta^{\alpha}_{\beta}\delta^{\nu}_{\mu})\,$. 
The operator $\Delta^{ex}_g$ commutes with all the other generators. 
As shown in~\cite{Boulanger:2004eh}, the gauge covariant algebra $\cal{G}$ is realized on
the space $\cal{W}$ of tensor fields $\cal{T}$ and generalized connections 
$\{\tilde{C}^N\}\,$. The second term on the right-hand side of (\ref{gcov3}) 
was not written in~\cite{Boulanger:2004eh}. However, it must be present in 
order for the commutation relation ${[}{\cal{D}}_{\beta},\mathbf{\Gamma}^{\alpha}{]}$ 
to be realized on the metric tensor as well, recalling
$\mathbf{\Gamma}^{\alpha}g_{\mu\nu} = 0 = {\cal{D}}_{\rho}g_{\mu\nu}\,$.  

The generalized connections $\{\tilde{C}^N\}$ present in (\ref{ansatz}) 
are obtained from \cite{Boulanger:2004eh}, 
setting the diffeomorphisms ghosts $\x^{\m}$ to zero. 
All of them are Grassmann-odd and read
\begin{eqnarray}
&\{\tilde{C}^N\} = \{ 2\omega\,,d{x}^{\nu}\,,
 \tilde{C}^{\nu}{}_{\mu}\,,\tilde{\omega}_{\alpha} \}\,,& 
\nonumber \\
&\tilde{C}^{\nu}{}_{\mu}=\Gamma_{~\,\mu\rho}^{\nu}\,d{x}^{\rho}\,,\quad
\tilde{\omega}_{\alpha}=\omega_{\alpha} - K_{\alpha\rho}\,d{x}^{\rho}\,, \quad
\omega_{\alpha}=\partial_{\alpha}\omega\,.&
\nonumber 
\end{eqnarray}
Then, with $\{\Delta_N\}=\{\Delta^{ex}_g \,, {\cal{D}}_{\nu}\,,{\Delta^{\mu}{}_{\nu}}\,,\mathbf{\Gamma}^{\alpha} \}\,$, 
the action of $\tilde{s}_{\!_W}$ on the tensor fields
$\{{\cal{T}}^i\}$ and generalized connections $\{\tilde{C}^N\}$ 
can be written in the very concise form
\begin{eqnarray}
	&\tilde{s}_{\!_W}{\cal{T}}^i = \tilde{C}^N \Delta_{N}{\cal{T}}^i\,,
	\quad
	\tilde{s}_{\!_W}\tilde{C}^N = \frac{1}{2}\,\tilde{C}^L\tilde{C}^K
	{\cal{F}}_{KL}^{~~~N}({\cal{T}})\,,&
	\nonumber
\end{eqnarray}
where ${\cal{F}}_{KL}^{~~~N}({\cal{T}})$ denote the structure functions
of the gauge covariant algebra $\cal{G}$: 
\begin{eqnarray}
&{[}\Delta_{M},\Delta_{N}{]}={\cal{F}}_{MN}^{~~~\;L}({\cal{T}})\Delta_{L}\,.& 
\nonumber
\end{eqnarray}
The relation $\tilde{s}_{\!_W}\tilde{C}^N = \frac{1}{2}\,\tilde{C}^L\tilde{C}^K
	{\cal{F}}_{KL}^{~~~N}({\cal{T}})\,$ generalizes the so-called ``Russian formula''. 
It is rather remarkable that the sole equations (\ref{sdg})--(\ref{sdw}) completely
determine the gauge covariant algebra (\ref{gcov1})--(\ref{algebra3}). 

A relevant issue concerning the algebra $\cal{G}$ given by (\ref{gcov1})--(\ref{algebra3})
(it is \emph{not} a Lie algebra) is whether it can be related to the conformal
algebra $\mathfrak{so}(n,2)\,$. After all, we are considering a general
class of theories that are classically diffeomorphism and Weyl invariant, 
and we know that such theories, in the flat limit 
$g_{\mu\nu}\rightarrow\eta_{\mu\nu}$, reduce to 
conformally-invariant theories. 
Introducing the new set of generators 
$\{\,P_{\mu}\,,\;K_{\nu}\,,\;M_{\mu\nu}\,,\;D\,\}$ via
\begin{eqnarray}
&\{\,\Delta_{\mu\nu}\,,\;\mathbf{\Gamma}_{\alpha}\,,\;D\,\}
=\{\,g_{\mu\rho}\Delta^{\rho}{}_{\nu}\,,\;g_{\alpha\beta}\mathbf{\Gamma}^{\beta}\,,\;
\delta^{\mu}_{\nu}\Delta^{\nu}{}_{\mu}-\Delta^{ex}_g \,\}\,,&
\nonumber \\
&\{\,P_{\mu}\,,\;K_{\nu}\,,\;M_{\mu\nu}\,\} = 
\{\,\frac{1}{4}\,{\cal{D}_{\mu}}\,,\;2\,\mathbf{\Gamma}_{\nu}\,,\;-2\,\Delta_{[\mu\nu]}\,\}
 	\,,& 
 	\nonumber
\end{eqnarray}
one gets from (\ref{gcov1})--(\ref{algebra3}) the following
gauge algebra:
\begin{eqnarray}
	{[}P_{\alpha},M_{\mu\nu}{]} &=& 2 \,g_{\alpha[\mu}P_{\nu]}\,,
	\quad 
	{[}K_{\alpha},M_{\mu\nu}{]} = 2 \,g_{\alpha[\mu}K_{\nu]}\,,
\nonumber \\
  {[}D,P_{\mu}{]} &=& P_{\mu}\,,
	\quad  {[}D,K_{\mu}{]} = -K_{\mu}\,,
\nonumber \\
	{[}M_{\alpha\mu},M_{\beta\nu}{]} 
	 &=& 2\, g_{\alpha[\beta}M_{\nu]\mu} - 2\, g_{\mu[\beta}M_{\nu]\alpha} \,,
\nonumber \\
	{[}P_{\mu},K_{\nu}{]} &=& 2\, (g_{\mu\nu}D + M_{\mu\nu})\,,\quad
	{[}K_{\mu},K_{\nu}{]} = 0\,,
\nonumber \\
  {[}P_{\mu},P_{\nu}{]} &=& -\frac{1}{2}\, W^{\rho\sigma}_{~~~\mu\nu}\,M_{\rho\sigma} 
  -\frac{1}{2}\,C_{\alpha\mu\nu} \,K^{\alpha} 
\nonumber 
\end{eqnarray}
which is isomorphic to the conformal algebra $\mathfrak{so}(n,2)$ when  
$g_{\mu\nu}=\eta_{\mu\nu}\,$, as was to be expected.  
Discussions and references on soft algebras, soft group manifolds and the transition from 
curved to flat spacetime in this context can be found in~\cite{Hehl:1994ue}. 

After this short comment on the relation between the soft (gauge) covariant algebra 
$\cal{G}$ and the (rigid) conformal algebra $\mathfrak{so}(n,2)$, 
we can proceed with the 
solution of the WZ consistency condition for the Weyl anomaly and 
its schematic solution (\ref{ansatz}). 
Because of the fermionic nature of the Weyl ghost $\omega$, 
the generalized connections $\tilde{C}^{N_i}$ in (\ref{ansatz}) must all
be different from $2\omega\,$, otherwise $\alpha({\cal{W}})$ vanishes. 
The appearance of the undifferentiated Weyl ghost $\omega$ in (\ref{ansatz}) 
is not an assumption. The Weyl-ghost dependence of the anomaly $a^n_1$ 
can entirely be expressed in terms of the undifferentiated ghost $\omega\,$, 
by integrating by parts:  
$\sqrt{-g}\;\omega_{\alpha}\,V^{\alpha}=\,$
$\partial_{\alpha}(\omega\,\sqrt{-g}\; V^{\alpha})- \,\omega \,\sqrt{-g}\;
\nabla_{\alpha}V^{\alpha}\,$.
We can now proceed with (\ref{cohoproblemweyl}) and expand $\alpha({\cal{W}})$ 
in powers of the connection $\tilde{C}^{\nu}{}_{\mu}\,$, 
\begin{eqnarray}
	\alpha({\cal{W}})&=&\sum_{k=0}^{m}\alpha({\cal{W}})\;,
	\quad N_C\,\alpha_k=k\,\alpha_k\;,
	\nonumber \\ 
	N_C &=& \tilde{C}^{\nu}{}_{\mu}\frac{\partial^L}{\partial \tilde{C}^{\nu}{}_{\mu}} \;.
\nonumber
\end{eqnarray}
On ${\cal{W}}\,$, the differential $\tilde{s}_{\!_W}$ decomposes into three parts, 
\begin{eqnarray}
	\tilde{s}_{\!_W}\, \alpha({\cal{W}}) = (\tilde{s}_{\!_W}^{lie} + \tilde{s}_{\!_W}^0 + \tilde{s}_{\!_W}^{-1})\,\alpha({\cal{W}})
\end{eqnarray}
which have $N_C$-degrees $1$, $0$, $-1$ respectively. 

The action of $\tilde{s}_{\!_W}^{lie}$, $\tilde{s}_{\!_W}^{0}$ and 
$\tilde{s}_{\!_W}^{-1}$ can
be summarized in Table \ref{ta}, together with
\begin{eqnarray}
	\tilde{s}_{\!_W}^{-1}\tilde{C}^{\nu}{}_{\mu}&=&\frac{1}{2}\,d{x}^{\rho} 
	d x^{\sigma}W_{~\,\mu\rho\sigma}^{\nu}+
  {\cal{P}}^{\nu\alpha}_{\beta\mu}\;\tilde{\omega}_{\alpha}\;d{x}^{\beta}\,.
\nonumber
\end{eqnarray} 
\begin{table}[h]
\begin{center}
\begin{tabular}{|c||c|c|c|}
\hline
     & $\tilde{s}_{\!_W}^{lie}$ &  $\tilde{s}_{\!_W}^{0}$  &  $\tilde{s}_{\!_W}^{-1}$ \\
\hline \hline
$\tilde{C}^{\nu}{}_{\mu}$ & $-\tilde{C}^{\nu}{}_{\alpha}\tilde{C}^{\alpha}{}_{\mu}$ & 
$0$ &  $\tilde{s}_{\!_W}^{-1}\tilde{C}^{\nu}{}_{\mu}$ \\
\hline
$\tilde{\omega}_{\alpha}$ & $\tilde{C}^{\beta}{}_{\alpha}\tilde{\omega}_{\beta}$ & 
$\frac{1}{2}\,d x^{\rho} d x^{\sigma} {C}_{\alpha\rho\sigma}$ & $0$ \\
\hline
$\omega$ & $0$ & $dx^{\mu} \tilde{\omega}_{\mu}$ & $0$ \\
\hline
$g_{\mu\nu}$ & $\tilde{C}^{\beta}{}_{\alpha}\Delta^{\alpha}{}_{\beta}g_{\mu\nu}$ & 
$2\,\omega \,g_{\mu\nu}$ & $0$ \\
\hline
$W_{\Omega_i}$ & $\tilde{C}^{\beta}{}_{\alpha}\Delta^{\alpha}{}_{\beta}W_{\Omega_i}$ & 
$dx^{\mu}{\cal{D}}_{\mu}W_{\Omega_i}+\tilde{\omega}_{\alpha} \mathbf{\Gamma}^{\alpha}
W_{\Omega_i}$ & $0$ \\
\hline
\end{tabular}
\caption{Decomposition of the action of $\tilde{s}_{\!_W}$\label{ta}}
\end{center}
\end{table}

The cocycle condition $\tilde{s}_{\!_W}\alpha=0\,$ thus decomposes into 
\begin{eqnarray}
	0 &=& \tilde{s}_{\!_W}^{lie} \alpha_m \label{1} \\
	0 &=& \tilde{s}_{\!_W}^{0} \alpha_m + \tilde{s}_{\!_W}^{lie} \alpha_{m-1} \label{2} \\
	0 &=& \tilde{s}_{\!_W}^{-1} \alpha_m + \tilde{s}_{\!_W}^{0} \alpha_{m-1} + 
	\tilde{s}_{\!_W}^{lie} \alpha_{m-2} 
	\nonumber 
	\\
	&\vdots&\nonumber
\end{eqnarray}
In the first equation, a contribution of the form $\tilde{s}_{\!_W}^{lie} \beta_{m-1}$ 
can be redefined away by subtracting the trivial piece $\tilde{s}_{\!_W}\beta_{m-1}$ 
from $\alpha\,$. The solution of equation (\ref{1}) is known because we know the Lie algebra 
cohomology of $\mathfrak{gl}(n)\,$.  
Indeed, $\mathfrak{gl}(n)\cong \mathbb{R}\oplus \mathfrak{sl}(n)\,$ is reductive.  
Since all the fields of $\cw$ transform according to finite-dimensional linear 
representations of $\mathfrak{gl}(n)$, we have
\begin{eqnarray}
	\alpha_m = \varphi_i(dx, \omega, \tilde{\omega}_{\alpha}, {\cal{T}}) 
	P^i(\tilde{\theta})\,, \quad
	\tilde{s}_{\!_W}^{lie} \varphi_i = 0\,.
	\label{sollie} 
\end{eqnarray}
The $P^i(\tilde{\theta})$ are linearly independent polynomials in the primitive elements 
$\tilde{\theta}_K$ of the Lie algebra cohomology of $\mathfrak{gl}(n)$. 
The $\tilde{\theta}_K$'s are monomials in the $\tilde{C}^{\nu}{}_{\mu}$'s 
and correspond to the independent Casimir operators of $\mathfrak{gl}(n)\,$.

Inserting (\ref{sollie}) in (\ref{2}) gives 
$$(\tilde{s}_{\!_W}^{0} \varphi_i)P^i(\tilde{\theta})+\tilde{s}_{\!_W}^{lie}\alpha_{m-1}=0\,.$$
Again, using the Lie algebra cohomology, we deduce 
\begin{eqnarray}
	\tilde{s}_{\!_W}^{0} \varphi_i(dx,\omega,\tilde{\omega}_{\alpha},{\cal{T}})
	=0\quad \forall\; i\,.
	\label{4}
\end{eqnarray}
We can assume that none of the $\varphi_i$'s is of the form 
$\tilde{s}_{\!_W}\vartheta(dx,\omega,\tilde{\omega}_{\alpha},{\cal{T}})$ 
because otherwise we could remove that particular $\varphi_i$ by subtracting the trivial piece 
$\tilde{s}_{\!_W}(\vartheta P^i)$ from $\alpha\,$. 
Such a subtraction does not clash with the other 
redefinitions made so far. In particular it does not reintroduce a term 
$\tilde{s}_{\!_W}^{lie}\beta_{m-1}$ in (\ref{sollie}) because of the definition of the $P^i$'s.    

Hence, since the $\varphi_i$'s do not depend on the $\tilde{C}^{\nu}{}_{\mu}$'s, 
we see that they are determined by the $\tilde{s}_{\!_W}$-cohomology in the space of 
$\mathfrak{gl}(n)$-invariant local total forms 
$\varphi(dx,\omega,\tilde{\omega}_{\alpha},{\cal{T}})$. 
[The coboundary condition 
$\varphi(dx,\omega,\tilde{\omega}_{\alpha},{\cal{T}})=\tilde{s}_{\!_W} \vartheta(dx,\omega,\tilde{\omega}_{\alpha},{\cal{T}})$
requires $\vartheta$ to be $\mathfrak{gl}(n)$-invariant, 
by expanding the equation in $\tilde{C}^{\nu}{}_{\mu}$.]
We thus have to solve 
\begin{eqnarray}
	&\tilde{s}_{\!_W} \varphi(dx,\omega,\tilde{\omega}_{\alpha},{\cal{T}})=0\,,&
	\label{equivcoho1} \\
	&\quad \varphi(dx,\omega,\tilde{\omega}_{\alpha},{\cal{T}})\neq 
	\tilde{s}_{\!_W} \vartheta(dx,\omega,\tilde{\omega}_{\alpha},\ct)\,,&
	\label{equivcoho2} \\ 
	&\tilde{s}_{\!_W}^{lie}\varphi = 0 = \tilde{s}_{\!_W}^{lie}\vartheta\,.&  
  \label{equivcoho3}
\end{eqnarray}

In order to solve the above equations, we decompose the relation
$\tilde{s}_{\!_W} \varphi(dx,\omega,\tilde{\omega}_{\alpha},{\cal{T}})=0$ 
into parts with definite degree in the appropriately symmetrized $W$-tensor 
fields (see \cite{Boulanger:2004eh}) 
and analyze it starting from the part with lowest degree. 
The decomposition is unique and thus well-defined thanks to the algebraic independence of 
the appropriately symmetrized $W$-tensors. The decomposition of 
$\tilde{s}_{\!_W}$ takes the form 
$\tilde{s}_{\!_W}=\sum_{k\geqslant 0}\tilde{s}_{\!_W}^{(k)}\,$, 
$[N_W,\tilde{s}_{\!_W}^{(k)}]=k\,\tilde{s}_{\!_W}^{(k)}$ 
where $N_W$ is the counting operator for the 
--- appropriately symmetrized --- $W$-tensors. 

The $\mathfrak{gl}(n)$-invariant local total form 
$\varphi(dx,\omega,\tilde{\omega}_{\alpha},{\cal{T}})$ 
decomposes into a sum of $\mathfrak{gl}(n)$-invariant terms 
\begin{eqnarray}
	\varphi(dx,\omega,\tilde{\omega}_{\alpha},{\cal{T}}) &=&
	\varphi_{(0)}(dx,\omega,\tilde{\omega}_{\alpha},g_{\mu\nu}) 
	\nonumber \\ 
	&&\quad\quad +\;
	\sum_{k>0}\varphi_{(k)}(dx,\omega,\tilde{\omega}_{\alpha},{\cal{T}})\,,
	\nonumber \\
	N_W\,\varphi_{(k)} &=& k \,\varphi_{(k)}\,.
	\nonumber
\end{eqnarray}  
The condition $\tilde{s}_{\!_W}\varphi=0$ requires, at lowest order in the tensor fields, 
\begin{eqnarray}
	\tilde{s}_{\!_W}^{(0)} \varphi_{(0)}(dx,\omega,\tilde{\omega}_{\alpha},g_{\mu\nu})=0\,.
	\label{cocyclegammazero}
\end{eqnarray}
Furthermore, we can remove any piece of 
the form 
$\tilde{s}_{\!_W}^{(0)} \vartheta_{(0)}(dx,\omega,\tilde{\omega}_{\alpha},g_{\mu\nu})$ 
from $\varphi_{(0)}$ by subtracting the trivial piece $\tilde{s}_{\!_W}\vartheta_{(0)}$ 
from $\varphi\,$. 
Hence, $\varphi_{(0)}$ is actually determined 
by the $\tilde{s}_{\!_W}^{(0)}$-cohomology in the space of $\mathfrak{gl}(n)$-invariant local total forms with no dependence on the $W$-tensors. 
In particular, we can assume 
$\varphi_{(0)}\neq \tilde{s}_{\!_W}^{(0)} \vartheta_{(0)}(dx,\omega,\tilde{\omega}_{\alpha},g_{\m\n})\,$.
Writing $\varphi_{(0)}=\omega \,\ell_{(0)}(dx,\tilde{\omega}_{\alpha},g_{\mu\nu})\,$, 
the condition (\ref{cocyclegammazero}) 
translates into $dx^{\m}\tilde{\omega}_{\m}\ell_{(0)}=0\,$. 
The most general $\ell_{(0)}(dx,\tilde{\omega}_{\alpha},g_{\mu\nu})$ reads
\begin{eqnarray}
&\ell_{(0)}(dx,\tilde{\omega}_{\alpha},g_{\mu\nu})=\sum_{p=0}^{n}\eta_{p}\,dx^{\alpha_1}\ldots dx^{\alpha_p}\,
\tilde{\omega}_{\alpha_1}\ldots	\tilde{\omega}_{\alpha_p}+&
\nonumber \\
&\frac{1}{\sqrt{-g}}\,\sum_{p=0}^n 
\lambda_p\,\varepsilon^{\nu_1\ldots\nu_p\mu_1\ldots\mu_{n-p}}\,
g_{\mu_1\alpha_1}\ldots g_{\mu_{n-p}\alpha_{n-p}}&
\nonumber \\
 &\times dx^{\alpha_1}\ldots dx^{\alpha_{n-p}}\,\tilde{\omega}_{\nu_1}\ldots	\tilde{\omega}_{\nu_p}\,,&
\nonumber
\end{eqnarray}
where $\eta_p$ and $\l_p\,$ are constants, $0\leqslant p\leqslant n\,$. 
In the second line of the above equation, we have inserted an appropriate power of 
$\det(g_{\mu\nu})$ in order that the corresponding local total form $\varphi$ 
possesses the correct weight to provide us with a candidate anomaly 
(the $\varepsilon$-symbol is the completely antisymmetric weight--$1$ Levi-Civita tensor density), 
as imposed by condition (\ref{coho1}).  
The condition $dx^{\mu}\tilde{\omega}_{\mu}\ell_{(0)}=0$ imposes $\eta_p=0\,$, 
$0\leqslant p\leqslant n-1\,$, which yields   
\begin{eqnarray}
&\varphi_{(0)}(dx,\tilde{\omega}_{\alpha},g_{\mu\nu})=
\eta_{n}\,\omega\,\,dx^{\alpha_1}\ldots dx^{\alpha_n}\,
\tilde{\omega}_{\alpha_1}\ldots	\tilde{\omega}_{\alpha_n}+&
\nonumber \\
&\frac{\omega}{\sqrt{-g}}\sum_{p=0}^n 
\lambda_p\,\varepsilon^{\nu_1\ldots\nu_p\mu_1\ldots\mu_{n-p}}\,
g_{\mu_1\alpha_1}\ldots g_{\mu_{n-p}\alpha_{n-p}}&
\nonumber \\
 &\times dx^{\alpha_1}\ldots dx^{\alpha_{n-p}}\,\tilde{\omega}_{\nu_1}\ldots	\tilde{\omega}_{\nu_p}\,.&
\nonumber
\end{eqnarray}
However, the first term is a local total form of degree $2n+1\,$, which is too much since 
we look for local total forms of degree $n+1$ \footnote{At most, the corresponding 
factor $P(\tilde{\theta})$ being in this case $P(\tilde{\theta})=1$ and the Weyl anomaly 
thus reducing to $\alpha= \alpha_m = \varphi$, cf. (\ref{sollie}).}. 
Accordingly, we set $\eta_n=0\,$. 
 
The next step consists in determining whether $\varphi_{(0)}$ is 
$\tilde{s}_{\!_W}^{(0)}$-trivial 
or not. We find that all the terms in $\varphi_{(0)}$ are 
$\tilde{s}_{\!_W}^{(0)}$-trivial, except one.
Indeed, 
\begin{eqnarray}
 \tilde{s}_{\!_W}^{(0)}&\Big(&\frac{1}{\sqrt{-g}}\;
 \varepsilon^{\nu_1\ldots\nu_p\mu_1\ldots\mu_{n-p}}\,
 g_{\mu_1\alpha_1}\ldots g_{\mu_{n-p}\alpha_{n-p}}
\nonumber \\
 &\times& 
 dx^{\alpha_1}\ldots dx^{\alpha_{n-p}}\,\tilde{\omega}_{\nu_1}\ldots\tilde{\omega}_{\nu_p} 
                    ~~\Big)
\nonumber \\
	&=&\omega\,\frac{(n-2p)}{\sqrt{-g}}\; \varepsilon^{\nu_1\ldots\nu_p\mu_1\ldots\mu_{n-p}}\,
  g_{\mu_1\alpha_1}\ldots g_{\mu_{n-p}\alpha_{n-p}}
\nonumber \\
  &&\times \; dx^{\alpha_1}\ldots dx^{\alpha_{n-p}}\,\tilde{\omega}_{\nu_1}
  \ldots\tilde{\omega}_{\nu_p}\;, 
\nonumber
\end{eqnarray}
so that only the term with $p=n/2\,$ survives in the $\tilde{s}_{\!_W}^{(0)}$-cohomology, leaving us with an $(n+1)$-total form $\varphi_{(0)}$.   

Summarizing, with $m=\frac{n}{2}$ we have (up to an irrelevant constant coefficient)
\begin{eqnarray}
	\varphi_{(0)} = \frac{\omega}{\sqrt{-g}}\,
	\varepsilon^{\nu_1\ldots\nu_m}_{\quad\quad~\mu_1\ldots\mu_m}\,
  dx^{\mu_1}\ldots dx^{\mu_m}\,\tilde{\omega}_{\nu_1}\ldots\tilde{\omega}_{\nu_m}.
  \label{fzero} 
\end{eqnarray}
Of course, this term exists only in even dimensions. 

We may now ask what is the completion $\varphi=\varphi_{(0)}+\sum_k \varphi_{(k)}$ 
of (\ref{fzero}) that would be invariant under the full differential 
$\tilde{s}_{\!_W}\,$. 
This question can be answered by using a decomposition of $\varphi$ and 
$\tilde{s}_{\!_W}$ with respect to the $\tilde{\omega}_{\alpha}$-degree.  
The differential $\tilde{s}_{\!_W}$ decomposes into a part noted 
$\tilde{s}_{\flat}$ which lowers the 
$\tilde{\omega}_{\alpha}$-degree by one unit, 
a part noted $\tilde{s}_{\natural}$ which does not 
change the $\tilde{\omega}_{\alpha}$-degree and a part noted 
$\tilde{s}_{\sharp}$ which raises the 
$\tilde{\omega}_{\alpha}$-degree by one unit:  
$\tilde{s}_{\!_W}= 
\tilde{s}_{\flat}+\tilde{s}_{\natural}+\tilde{s}_{\sharp}$.
The action of these three parts of $\tilde{s}_{\!_W}$ is given in Table \ref{ta2}. 
\begin{table}[h]
\begin{center}
\begin{tabular}{|c||c|c|c|}
\hline
     & $\tilde{s}_{\flat}$ &  $\tilde{s}_{\natural}$  &  
     $\tilde{s}_{\sharp}$ \\
\hline \hline
$\tilde{\omega}_{\alpha}$ & 
$\frac{1}{2}\,d x^{\rho} d x^{\sigma} {C}_{\alpha\rho\sigma}$ & 
$\tilde{C}^{\beta}{}_{\alpha} \tilde{\omega}_{\beta}$ & $0$ \\
\hline
$\omega$ & $0$ & $ 0 $ & $dx^{\mu} \tilde{\omega}_{\mu}$ \\
\hline
$W_{\Omega_i}$ & $ 0 $ & 
$\tilde{C}^{\beta}{}_{\alpha}\Delta^{\alpha}{}_{\beta}W_{\Omega_i}
+dx^{\mu}{\cal{D}}_{\mu}W_{\Omega_i}$ & 
$\tilde{\omega}_{\alpha} \mathbf{\Gamma}^{\alpha}W_{\Omega_i}$ \\
\hline
$g_{\mu\nu}$ & $ 0 $ & 
$\tilde{C}^{\beta}{}_{\alpha}\Delta^{\alpha}{}_{\beta}\,g_{\mu\nu}
+2\,\omega\, g_{\mu\nu}$ & $0$ \\
\hline
$\tilde{C}^{\nu}{}_{\mu}$ & 0 & 
$-\tilde{C}^{\nu}{}_{\alpha}\tilde{C}^{\alpha}{}_{\mu}+\frac{1}{2}\,d{x}^{\rho} 
d x^{\sigma}W_{~\,\mu\rho\sigma}^{\nu}$ & 
  ${\cal{P}}^{\nu\alpha}_{\beta\mu}\;\tilde{\omega}_{\alpha}\;d{x}^{\beta}$ \\
\hline
\end{tabular}
\caption{Action of $\tilde{s}_{\!_W}$, decomposed w.r.t the $\tilde{\omega}_{\alpha}$-degree \label{ta2}}
\end{center}
\end{table}

The decomposition of $\varphi$ with respect to the $\tilde{\omega}_{\alpha}$-degree 
reads
\begin{eqnarray}
	\varphi&=&\Phi^{[m]}_m + \Phi^{[m+1]}_{m-1}+\ldots +\Phi^{[n-1]}_1 + \Phi^{[n]}_0\,, 
\nonumber \\
  &&\qquad \Phi^{[m]}_m = \varphi_{(0)}\,,\qquad m=\frac{n}{2}\;, 
\nonumber
\end{eqnarray}
where each term $\Phi^{[n-r]}_r\,$ ($0\leqslant r\leqslant m$) is $\mathfrak{gl}(n)$-invariant, 
possesses a $\tilde{\omega}_{\alpha}$-degree $r\,$ and explicitly contains the 
product of $(n-r)$ $dx$'s. [Of course, some $dx$'s are also hidden inside 
the $\tilde{\omega}_{\a}$'s.]   

Decomposing the cocycle condition $\tilde{s}_{\!_W} \varphi = 0 $ with respect to the $\tilde{\omega}_{\alpha}$-degree 
yields the following descent of equations
\begin{eqnarray}
   \tilde{s}_{\flat}\Phi^{[n-1]}_1 + \tilde{s}_{\natural} \Phi^{[n]}_0 &=& 0\quad,
\nonumber \\
	\tilde{s}_{\flat}\Phi^{[n-2]}_2 + \tilde{s}_{\natural}\Phi^{[n-1]}_1 + \tilde{s}_{\sharp}\Phi^{[n]}_0 &=& 0 \quad ,
\nonumber \\
            &\vdots&
\nonumber \\
	\tilde{s}_{\flat}\Phi^{[m]}_m + \tilde{s}_{\natural}\Phi^{[m+1]}_{m-1} 
	+ \tilde{s}_{\sharp}\Phi^{[m+2]}_{m-2} &=& 0 \quad ,
\nonumber \\            
  \tilde{s}_{\natural}\Phi^{[m]}_m + \tilde{s}_{\sharp}\Phi^{[m+1]}_{m-1} &=& 0 \quad ,
\nonumber \\
 \tilde{s}_{\sharp}\Phi^{[m]}_m &=& 0 \quad .
\nonumber
\end{eqnarray}

In the following theorem, we give the expression for $\Phi^{[n-r]}_r\,$, 
$0\leqslant r\leqslant m\,$, such that $\varphi=\sum_{r=0}^{m}\Phi^{[n-r]}_r$ 
is a solution of $\tilde{s}_{\!_W} \varphi=0\,$ with 
$\Phi^{[m]}_m = \varphi_{(0)}$ (\ref{fzero}).
Furthermore, the $n$-form $\Phi^{[n]}_0\,$ is separately 
 $\tilde{s}_{\!_W}$-invariant and the top form degree component of  
 $\varphi$ 
 is nothing but the type\,-A Weyl anomaly. 
 The anomaly $\beta=\Phi^{[n]}_0$ gives rise to a trivial descent and is a linear 
 combination of type\,-B anomalies obtained simply by contractions of products of Weyl tensors.

\vspace*{.3cm}

{\underline{\it{Theorem 1\,:}}} ~
Let $\psi_{\m_1\ldots\m_{2p}}$ be the local total form 
\begin{eqnarray}
	\psi_{\mu_1\ldots\mu_{2p}} &=& 
	\frac{\omega}{\sqrt{-g}} \;
	\varepsilon^{\alpha_1\ldots\alpha_r}_{\quad\quad~\,\nu_1\ldots\nu_r\mu_1\ldots\mu_{2p}}
	\nonumber \\
	&& \qquad\qquad \times ~ \tilde{\omega}_{\alpha_1}\ldots\tilde{\omega}_{\alpha_r}
	\;d x^{\nu_1}\ldots d x^{\nu_r}\,,
	\nonumber \\
	p&=&m-r\,,\quad m=n/2\,,\quad 0\leqslant r\leqslant m\,
  \nonumber
\end{eqnarray}
and $W^{\mu\nu}$ the tensor-valued two-form
\begin{eqnarray}
	W^{\mu\nu} &=& W^{\mu}_{~\;\lambda}\,g^{\lambda\nu}=
	\frac{1}{2}\,d{x}^{\rho} d x^{\sigma}W_{~\,\lambda\rho\sigma}^{\mu}\,g^{\lambda\nu}\,.
\nonumber
\end{eqnarray}
Then, the local total forms ${\Phi}_r^{[n-r]}$ $(0 \leqslant r \leqslant m)$
\begin{eqnarray}
	\Phi^{[n-r]}_{r} = \frac{(-1)^p}{2^p}\,\frac{m!}{r!\,p!}\;\psi_{\m_1\ldots\m_{2p}}\,
	W^{\m_1\m_2}\ldots \,W^{\m_{2p-\!1}\m_{2p}}
\nonumber
\end{eqnarray}
obey the descent of equations
\begin{eqnarray}
&& \left\{
\begin{array}{cl}
	\tilde{s}_{\flat}\Phi^{[n-r]}_{r} + \tilde{s}_{\natural}\Phi^{[n-r+1]}_{r-1} &=\; 0\quad,
	\nonumber \\
	\tilde{s}_{\sharp}\Phi^{[n-r]}_{r} &=\; 0\quad,\quad (1\leqslant r\leqslant m)
	\nonumber 
\end{array}\right.
\\
  &&\tilde{s}_{\flat}\Phi^{[n-1]}_{1} \;=\; 0 \;=\; \tilde{s}_{\!_W} \Phi^{[n]}_{0}\;,	
\nonumber
\end{eqnarray}
so that the following relations hold: 
\begin{eqnarray}
	\tilde{s}_{\!_W} {\alpha} \,= \!&0&\! =\; \tilde{s}_{\!_W}{\beta} \;,
	\nonumber\\
	{\alpha}&=&\sum_{r=1}^{m}\Phi^{[n-r]}_{r} \;, \quad {\beta} = \Phi^{[n]}_{0}\,. 
\nonumber
\end{eqnarray}
\vspace*{.1cm}

{\underline{\it{Proof\,:}}} ~ The proof follows by direct computation, 
using the tracelessness of the Weyl tensor and with the help of the identity 
$\nabla W^{\mu\nu}=2\,{C}_{\rho}\,g^{\rho[\mu}dx^{\nu]}\,$
relating the covariant differential of the Weyl two-form $W^{\mu\nu}$
to the Cotton two-form 
${C}_{\rho}=\frac{1}{2}\,dx^{\mu}dx^{\nu}\,{C}_{\rho\mu\nu}$. 

\vspace*{.3cm}

\noindent Finally, we have the
\vspace*{.3cm}

{\underline{\it{Theorem 2\,:}}} ~
(A) The top form-degree component $a^n_1$ of ${\alpha}$ (cf. Theorem 1) 
satisfies the WZ consistency conditions for the Weyl anomalies. 
The WZ conditions for $a^n_1$ give rise to a non-trivial descent 
and $a^n_1$ is the \textit{unique} anomaly with such a property,  
up to the addition of trivial terms and anomalies satisfying a trivial descent. 

(B) The anomaly $\beta=\Phi^{[n]}_{0}$ satisfies a trivial descent and is 
obtained by taking contractions of products of Weyl tensors 
($m$ of them in dimension $n=2m$). 
The top form-degree component $e^n_1$ of $(\alpha+\beta)$
is proportional to the Euler density of the manifold ${\cal{M}}_{n}\,$: 
\begin{eqnarray}
e^n_1 \;=\;\frac{(-1)^m}{2^m}\; 
\omega \;(R_{a_1 b_1}\wedge\ldots\wedge R_{a_m b_m})\;
\varepsilon^{a_1b_1\ldots \,a_m b_m}\;.
\nonumber  	
\end{eqnarray}

\vspace*{.2cm}
{\underline{{\it{Proof\,:}}} ~ 
\vspace*{.2cm}

(A) When computing the solutions of (\ref{equivcoho1})--(\ref{equivcoho3}), 
we used an expansion of $\varphi(dx,\omega,\tilde{\omega}_{\alpha},{\cal{T}})$ 
in the number of 
(appropriately symmetrized) $W$-tensors and found a solution starting with a 
$W$-independent term $\varphi_{(0)}\,$ given in (\ref{fzero}). This term, as
we showed, gives rise to (a representative of) the so-called type\,-A anomaly. 
However, in order to compute the general solutions of 
(\ref{equivcoho1})--(\ref{equivcoho3}), we must determine whether other
solutions exist, that would start with a term $\varphi_{(\ell)}$ with $\ell>0\,$. 
If one returns to the decomposition of local total forms in terms of 
form degree and ghost number, writing 
$\varphi(dx,\omega,\tilde{\omega}_{\alpha},{\cal{T}})=\sum_{r=1}^{q+1} b_{r}^{p-r+1}$, 
the problem (\ref{equivcoho1})--(\ref{equivcoho3}) 
takes on the usual descent-equation form
\begin{eqnarray}
	s_{\!_W} b^p_1 + d \,b^{p-1}_2 &=& 0\quad,
	\label{ty} \\
	s_{\!_W} b^{p-1}_2 + d \,b^{p-2}_3 &=& 0\quad,
  \nonumber \\
	&\vdots&
	\nonumber \\
	s_{\!_W} b^{p-q+1}_q + d \,b^{p-q}_{q+1} &=& 0\quad,
\label{sec}\\
	s_{\!_W} b^{p-q}_{q+1}  &=& 0\quad (0\leqslant q\leqslant p \leqslant n), 
\label{first}	
\end{eqnarray}
where every element $b^{p-i}_{i+1}$
$(0\leqslant i\leqslant q)$ transforms as a local $(p-i)$-form under 
spacetime diffeomorphisms, so that $d \,b^{p-i}_{i+1}=\nabla b^{p-i}_{i+1}$
where $\nabla=dx^{\mu}\nabla_{\mu}$ is the Levi-Civita covariant differential.  
One assumes that the descent is displayed in its shortest expansion, \textit{i.e.}
that $q$ is minimal. This means that $b^{p-q}_{q+1}$ is non-trivial
in $H^{q+1,p-q}(s_{\!_W}\vert d)$ since otherwise 
$b^{p-q}_{q+1}=s_{\!_W} \mu^{p-q}_{q} + d \,\mu^{p-q-1}_{q+1}$ and 
(\ref{sec}) would then become $s_{\!_W} [ b^{p-q+1}_q - d \mu^{p-q}_{q}] = 0$, 
which, upon redefining $b^{p-q+1}_q$, 
would imply that the descent has shortened by one step, 
contrary to the shortest-descent hypothesis.
 
A priori, the head of the descent, $b^p_1\,$, possesses a form degree $p\leqslant n$ 
because candidate anomalies are obtained by completing 
[see Eqs. (\ref{1})--(\ref{sollie})] 
the product $\varphi(dx,\omega,\tilde{\omega}_{\alpha},{\cal{T}})P(\tilde{\theta})$,   
where $P(\tilde{\theta})$ is a polynomial in the primitive elements 
$\tilde{\theta}_K$ of the Lie algebra cohomology of $\mathfrak{gl}(n)\,$ and possesses
a non-vanishing form degree, except for the trivial element $P(\tilde{\theta})=1\,$.
The ghost number of $b^p_1$ must be one because the 
$P^i(\tilde{\theta})$'s have a vanishing Weyl-ghost degree. 
On the other hand, it is known that the condition (\ref{coho1}), 
in the absence of (derivatives of) diffeomorphisms ghosts, 
admits only two kinds of terms~\cite{Brandt:1989et}.  
The first have the general form ${\cal{L}}d^nx$ where the lagrangian density 
$\cal{L}$ is constructed out of the Riemann tensor, the matter fields, 
the Yang-Mills field strength and their covariant derivatives. 
The second class of terms contains the pure-gravity Chern-Simons densities 
that depend explicitly on the Riemannian connection one-form $\tilde{C}^{\nu}{}_{\mu}$ 
and on the undifferentiated curvature two-form 
$R^{\mu}_{~\nu}=\frac{1}{2}\,R^{\mu}_{~\,\nu\rho\sigma}dx^{\rho}dx^{\sigma}\,$. 
Since the candidate
Weyl-anomalies are linear in the Weyl-ghost $\omega$ which plays the r\^ole
of a matter field, we conclude that no Chern-Simons term can appear in $a^n_1$, 
and hence the only allowed polynomial $P(\tilde{\theta})$ is the trivial
one, $P(\tilde{\theta})=1\,$, which in turn implies that one can set
$p=n$ in the descent (\ref{ty})--(\ref{first}), without loss of generality. 

The case where $q=0$ means that the descent is trivial and the candidate
anomalies satisfy $s_{\!_W} a^n_1 = 0\,$. These are the type\,-B Weyl anomalies
that can be classified and computed systematically along the lines of
\cite{Boulanger:2004zf,Boulanger:2004eh}. Accordingly, in what follows we assume $q>0$. 

The bottom of the descent is obtained from  
$\alpha({\cal{W}})$ by taking its maximal $\tilde{\omega}_{\alpha}$-degree 
component and taking only the contribution $\omega_{\alpha}$
of $\tilde{\omega}_{\alpha}=\omega_{\alpha}-dx^{\mu}K_{\mu\alpha}$. 
In other words, the bottom of the descent must not depend on 
the one-form potential ${\cal{A}}_{\alpha}=-dx^{\mu}K_{\mu\alpha}\,$.   
A priori, when determining the most general non-trivial bottom
$b^{n-q}_{q+1}$ in (\ref{first}), the dependence on the 
space of $W$-tensors can be complicated. 
However, it was proved in~\cite{Barkallil:2002fp} that, for
any given (super) Lie algebra ${\mathfrak{g}}$, the solutions of non-trivial descents as in 
(\ref{ty})--(\ref{first}) can be computed, without loss
of generality, in the small algebra ${\cal{B}}$ generated by the one-form potentials, 
the curvature two-forms, the ghosts and the exterior derivatives of the ghosts. 

In the present setting, the curvature two-forms decompose into
$W^{\mu}_{~\,\nu}=\frac{1}{2}\,d x^{\rho}d x^{\sigma}\,W^{\mu}_{~\;\nu\rho\sigma}$ and 
${C}_{\alpha}=$
$\frac{1}{2}\,d x^{\rho}d x^{\sigma}\,{C}_{\alpha\rho\sigma}$, which take 
their values along the generators $\Delta^{\nu}{}_{\mu}$ and $\mathbf{\Gamma}^{\alpha}$, 
respectively, as can be read off from~(\ref{algebra3}). 
The algebra generated by 
$\{\Delta^{\nu}{}_{\mu}, \mathbf{\Gamma}^{\a}\}$ [see (\ref{gcov1}), (\ref{gcov2})] 
is non-reductive, being isomorphic to the semi-direct sum of $\mathfrak{gl}(n)$ and 
the abelian translation-like algebra $\mathfrak{t}(n)\,$.
In analogy with a Yang-Mills gauge theory, the r\^ole of the Killing 
metric is played here by $g_{\m\nu}$ which obeys $\cd_{\rho}g_{\mu\nu}=0\,$. 
Another invariant object at our disposal is the Levi-Civita $\varepsilon$ symbol. 
The exterior differentials of the ghosts give $dx^{\alpha}\omega_{\alpha}$ and $dx^{\beta}\partial_{\beta}\omega_{\alpha}$, but the latter
must be rejected because they do not belong to ${\cal{W}}\,$.  

To summarize, the bottom of the descent $b^{n-q}_{q+1}$ can depend on 
the $W$-tensors only through the curvature two-forms ${C}_{\alpha}$ and 
${W}^{\mu}_{\,~\nu}$. It is linear in the undifferentiated ghost $\omega$ and must 
not depend on ${\cal{A}}_{\alpha}=-dx^{\mu}K_{\mu\alpha}\,$. 
Moreover, it is easy to see that the Cotton two-form
${C}_{\alpha}$ cannot enter $b^{n-q}_{q+1}$ since otherwise, 
up to a trivial $d$-exact term, $b^{n-q}_{q+1}$ would depend on ${\cal{A}}_{\alpha}\,$. 
This is because ${C}_{\alpha\mu\nu} = {2}\,\nabla_{[\nu}K_{\mu]\alpha}\,$ and the
fact that $\nabla$ may be replaced by the exterior differential $d$ inside the
descent made of $p\,$-forms. 

Hence, the general form of $b^{n-q}_{q+1}$ is
given by a linear combinaison of terms of the form 
$\omega \,\mbox{Tr}\,(\prod_{i,j,k}{W}^{\mu_i}_{~\;\nu_i}\,\omega_{\rho_j}dx^{\sigma_k})$ 
where the trace is obtained by using the metric and the $\varepsilon$ symbol.
The relation $W^{\mu}_{~\,\nu}\omega_{\mu}=-s_{\!_W} {C}_{\nu}$ 
(see e.g. \cite{Boulanger:2004eh})
shows that no $W^{\mu_i}_{~\;\nu_i}$ can be contracted with a $\omega_{\rho}$. 
Together with the identity ${W}^{\mu}_{~\;\nu}dx^{\rho}g_{\rho\mu}= 0$, 
this shows that
the indices of the ${W}^{\mu_i}_{~\;\n_i}$'s must be contracted among themselves. 
  
Suppose first that we use no Levi-Civita $\ve$ symbol in order to contract the indices 
in $\prod_{j,k}\omega_{\rho_j}dx^{\sigma_k}$.  
The corresponding $b^{n-q}_{q+1}$'s look like 
$b^{n-q}_{q+1}\sim \omega \,\mbox{Tr}\,(\prod_{i}{W}^{\mu_i}_{~\;\nu_i})\prod_j^q 
\omega_{\rho_j}dx^{\rho_j}$. 
Taking the exterior derivative of such a term gives contributions where $d$ hits
$\omega$ and contributions when $d$ hits one of the ${W}^{\mu_i}_{~\;\nu_i}$'s. 
Trivially, $d(\omega_{\alpha} dx^{\alpha})=0$ because 
$\omega_{\alpha}=\partial_{\alpha}\omega\,$. 
Because in $d \,b^{n-q}_{q+1}$ one can replace $d{W}^{\mu}_{~\,\nu}$ by  
$2\,{C}_{\rho}\,g^{\rho[\mu}dx^{\sigma]}\,g_{\sigma\nu}$
and because
${W}^{\mu}_{~\,\nu}dx^{\rho}g_{\rho\mu} = 0$, only the contribution from 
$d\,\omega$ survives in $d \,b^{p-q}_{q+1}\,$. 
This provides terms of the form 
$d \,b^{p-q}_{q+1}\sim  \,\mbox{Tr}\,(\prod_{i}{W}^{\mu_i}_{~\;\nu_i})\prod_j^{q+1} 
\omega_{\rho_j}dx^{\rho_j}$ that, in the space ${\cal{Y}}$ obtained from ${\cal{W}}$ 
by discarding the $\tilde{C}^{\mu}{}_{\nu}$'s, clearly
belong to the cohomology of $s_{\!_W}$ --- it suffices to use the results of~\cite{Boulanger:2001he}, taking the linearized part of $d \,b^{p-q}_{q+1}$  --- 
and therefore are obstructions to the lift (\ref{sec}) of $b^{n-q}_{q+1}$. 

The only other possibilities in the expression 
of the candidate $b^{n-q}_{q+1}$ are exhausted by
\begin{eqnarray}
b^{n-q}_{q+1}&\sim& \omega \,\mbox{Tr}\,(\prod_{i}{W}^{\mu_i}_{~\;\nu_i})\;
\sqrt{-g}\;\varepsilon_{\sigma_1\ldots\sigma_q\rho_1\ldots\rho_{n-q}}
\nonumber \\
&&\times ~
g^{\sigma_1\tau_1}\ldots g^{\sigma_{q}\tau_q}\;\omega_{\tau_1}\ldots\,\omega_{\tau_q}\,dx^{\rho_1}\ldots\, dx^{\rho_{n-q}}\;.
\nonumber
\end{eqnarray}
However, such terms are non-trivial in $H(s_{\!_W},{\cal{Y}})$ iff $q=n/2\,$. 
Since the factor $\mbox{Tr}\,(\prod_{i=1}^k {W}^{\mu_i}_{~\;\nu_i})$ brings
a form degree $2k$ and because the remaining factor in $b^{n-q}_{q+1}$
already gives an $n$-form
at the top of the descent, we conclude that $k=0$ and the bottom of the descent 
reduces to the only term $(m=n/2)$
\begin{eqnarray}
b^{m}_{m+1} &=& \omega\,\sqrt{-g}\;\varepsilon_{\sigma_1\ldots\sigma_m\rho_1\ldots\rho_{m}}\;
g^{\sigma_1\tau_1}\ldots g^{\sigma_{m}\tau_m}\;
\nonumber \\
&&\quad\times\;\omega_{\tau_1}\ldots\,\omega_{\tau_m}\,dx^{\rho_1}\ldots\, dx^{\rho_{m}}
 \label{fre}
\end{eqnarray}
which is contained in (\ref{fzero}). The latter term gives rise to the
candidate anomaly $\alpha$ presented in Theorem 1. Because (\ref{fre}) 
is non-trivial in the cohomology $H(s_{\!_W},{\cal{Y}})$, 
so is the corresponding $a^n_1$ in $H(s_{\!_W}\vert d)$, taking into
account (\ref{coho1}) and (\ref{coho3}).  
This proves part (A) of the theorem. 

Part (B) is proved by direct computation. 

\section{Conclusions and dicussions}

It was questioned in the introduction of~\cite{Bonora:1985cq} whether 
a general algorithm as in the case of the chiral anomalies 
could exist for the Weyl anomalies. 
Thanks to the efforts of many people, 
extending the cohomological method to arbitrary dimension has become 
doable, even for the Weyl anomalies. As we showed in the present paper, 
an algorithm as in the case of the non-Abelian chiral anomalies 
does indeed exist for the Weyl anomalies. It features descent equations 
\`a la Stora-Zumino and provides a general, purely algebraic understanding of the 
structure of the Weyl anomalies in arbitrary dimensions, thereby answering 
a question raised by Deser and Schwimmer~\cite{Deser:1993yx}.   

The approach followed here is purely cohomological  
and independent of any regularization scheme. 
No dimensional argument is used and the evenness of 
the spacetime dimension is a consequence of the Wess-Zumino consistency 
condition, as is the general structure of the Weyl anomalies.  

\section{acknowledgments}

This work was supported by the Fonds de la Recherche Scientifique, FNRS (Belgium). 

We thank G. Barnich for many useful discussions and M. Henneaux for having
suggested the project. We thank H. Osborn and Ph. Spindel
for their comments and encouragements.

\end{document}